\documentclass[12pt,journal, comsoc]{IEEEtran}

\ifCLASSINFOpdf
    \usepackage[pdftex]{graphicx}
    \graphicspath{{figures/}}
    \DeclareGraphicsExtensions{.pdf,.jpeg,.png, .eps}
\else
\fi

\usepackage{times}
\usepackage[utf8]{inputenc}
\usepackage[T1]{fontenc}
\usepackage[nolist]{acronym}
\usepackage[colorlinks=true,allcolors=blue]{hyperref}
\usepackage{url}
\usepackage{pifont}
\usepackage{multirow}
\usepackage{siunitx}
\usepackage{array,booktabs}

\usepackage[caption=false]{subfig}
\usepackage{tabularx}

\usepackage[usenames,dvipsnames,svgnames,table]{xcolor}
\usepackage[utf8]{inputenc}
\usepackage{tikz}
\usetikzlibrary{decorations,arrows,shapes}
\usepackage{amsmath}
\usepackage{color,soul}

\newcommand\copyrighttext{%
    \footnotesize The final publication is available at IEEE Communications Magazine  via \url{https://doi.org/10.1109/MCOM.001.2000877}}
    \newcommand\copyrightnotice{%
    \begin{tikzpicture}[remember picture,overlay]
    \node[anchor=south,yshift=10pt] at (current page.south) {\fbox{\parbox{\dimexpr\textwidth-\fboxsep-\fboxrule\relax}{\copyrighttext}}};
    \end{tikzpicture}%
}

\newcommand{\eg}{e.g., }

\newcommand{\ie}{i.e., }

\definecolor{gray}{gray}{0.9}
\definecolor{white}{gray}{1.0}
\newcolumntype{g}{>{\columncolor{gray}}l}
\newcolumntype{x}{>{\columncolor{gray}}p{1cm}}
\newcolumntype{y}{>{\columncolor{gray}}p{4cm}}
\newcolumntype{z}{>{\columncolor{gray}}p{2cm}}

\begin{acronym}
    \acro{pep}[PEP]{Performance Enhancing Proxy}
    \acro{loops}[LOOPS]{Local Optimizations on Path Segments}
    \acro{tsvwg}[TSVWG]{Transport Area Working Group}
    \acro{iccrg}[ICCRG]{Internet Congestion Control Research Group}
    \acro{rto}[RTO]{Retransmission timeout}
    \acro{fec}[FEC]{Forward Error Correction}
    \acro{holb}[HOL blocking]{Head-of-line blocking}
    \acro{ce}[CE]{Congestion Experienced}
    \acro{ecn}[ECN]{Explicit Congestion Notification}
    \acro{geneve}[Geneve]{Generic Network Virtualization Encapsulation}
    \acro{vxlan}[VXLAN]{Virtual Extensible LAN}
    \acro{nvgre}[NVGRE]{Network Virtualization using Generic Routing Encapsulation}
    \acro{stt}[STT]{Stateless Transport Tunneling}
    \acro{iesg}[IESG]{Internet Engineering Steering Group}
    \acro{vpn}[VPN]{Virtual Private Network}
    \acro{masque}[MASQUE]{Multiplexed Application Substrate over QUIC Encryption}
    \acro{socks}[SOCKS]{Socket Secure}
    \acro{sctp}[SCTP]{Stream Control Transmission Protocol}
    \acro{ice}[ICE]{Interactive Connectivity Establishment}
    \acro{taps}[TAPS]{Transport Services}
    \acro{dccp}[DCCP]{Datagram Congestion Control Protocol}
    \acro{icmp}[ICMP]{Internet Control Message Protocol}
    \acro{w3c}[W3C]{World Wide Web Consortium}
    \acro{grease}[GREASE]{Generate Random Extensions And Sustain Extensibility}
    \acro{ietf}[IETF]{Internet Engineering Task Force}
    \acro{bof}[BOF]{Birds of a Feather}
    \acro{wg}[WG]{Working Group}
    \acro{mptcp}[MPTCP]{Multipath TCP}
    \acro{bgp}[BGP]{Border Gateway Protocol}
    \acro{tcpm}[TCPM]{TCP Maintenance and Minor Extensions}
    \acro{irtf}[IRTF]{Internet Research Task Force}
    \acro{tcp}[TCP]{Transmission Control Protocol}
    \acro{bbr}[BBR]{Bottleneck Bandwidth and Round-trip propagation time}
    \acro{atsss}[ATSSS]{Access Traffic Steering, Switching, and Splitting}
    \acro{3gpp}[3GPP]{3rd Generation Partnership Project}
    \acro{cmt-sctp}[CMT-SCTP]{Concurrent Multipath Transfer SCTP}

\end{acronym}

\acrodefplural{vpns}[VPNs]{Virtual Private Networks}

\begin{document}

\title{Beyond QUIC v1 – A First Look at Recent Transport Layer IETF Standardization Efforts}
\author{Mike Kosek, Tanya Shreedhar, and Vaibhav Bajpai}

\maketitle

\begin{abstract} The transport layer is ossified.
    With most of the research and deployment efforts in the past decade focussing on the \ac{tcp} and its extensions, the QUIC standardization by the \ac{ietf} is to be finalized in early 2021.
    In addition to addressing the most urgent issues of TCP, QUIC ensures its future extendibility and is destined to drastically change the transport protocol landscape.
    In this work, we present a first look at emerging protocols and their \ac{ietf} standardization efforts \textit{beyond} QUIC \textit{v1}.
    While multiple proposed extensions improve on QUIC itself, \textit{\ac{masque}} as well as \textit{WebTransport} present different approaches to address long-standing problems, and their interplay extends on QUIC's take to address transport layer ossification challenges.
\end{abstract}

\begin{IEEEkeywords}
    Transport protocols, IETF, QUIC, MASQUE, WebTransport
\end{IEEEkeywords}
\vspace{-1em}

\copyrightnotice

\IEEEpeerreviewmaketitle

\section{Introduction} \label{sec:intro}

    The transport layer, which is responsible for the end-to-end connectivity between peers, is ossified \cite{deossifying}.
    While most of the transport protocol related research in the past decade focussed on the extension of TCP as the predominant transport protocol in the Internet, multiple studies showed that this quest is cumbersome, leading to slow adoption of innovative improvements like \ac{ecn} (RFC 3168) or \ac{mptcp} (RFC 6824), or even no adoption at all.
    Where previous efforts to deploy novel transport protocols like \ac{sctp} (RFC 4960) also did not lead to a wide adoption due to deployment obstacles for non-TCP/UDP-based protocols, using UDP as a substrate protocol promises Internet-scale deployability.
    Recently, QUIC as a UDP-based protocol set out to solve TCP's issues while ensuring its future extendibility \cite{quicinnovation}.
While QUIC is destined to drastically change the transport protocol landscape, TCP is still the most used protocol, and its importance will not diminish in the near future.
    Acting as a fallback protocol, services will be offered with both TCP \textit{and} QUIC for an extensive amount of time, and more specialized use-cases like \ac{bgp} (RFC 4271) might continue to use TCP indefinitely.
    To ensure continuous improvements, the \ac{ietf} \ac{tcpm} \ac{wg} discusses TCP and \ac{mptcp} modifications, and directs the standardization process for proposed specifications.
    Within \ac{tcpm}, RFC 8803 as well as RFC 8961 were recently standardized.
    The \textit{0-RTT TCP Convert Protocol} (RFC 8803) aims at improving the deployment of TCP extensions, where \textit{Requirements for Time-Based Loss Detection} (RFC 8961) discusses best practices to parameterize loss detection algorithms.
    \textit{The RACK-TLP loss detection algorithm for TCP} \cite{racktlp} was submitted to the \ac{iesg} for publication in December 2020, leveraging Recent Acknowledgements for fast recovery and improving on tail loss by explicitly triggering ACK feedback through Tail Loss Probes.
    While these additions are essential improvements to TCP, they may not overcome TCP's inherent issues.

    With the standardization of QUIC by the \ac{ietf} to be finalized in early 2021, QUIC's first version \textit{v1} (see §\ref{sec:quicv1}) addresses the most urgent issues of TCP such as multiplexing, \ac{holb}, mandatory encryption, as well as reduced connection establishment time with 0-RTT support while focussing on the web use-case, \ie delivery of web content to browsers.
    Extending on \textit{v1}, the \ac{wg} actively discusses future extensions, which we will detail in §\ref{quic}.
These extensions introduce improvements to version negotiation as well as connection IDs, add multipath capabilities, enable unreliable delivery within QUIC as well as HTTP/3, further extend the future useability of the QUIC protocol, and add performance improvements by negotiating acknowledgement handling.

    QUIC's mandatory encryption does provide challenges for specialized use-cases where end-to-end connectivity is not possible (\eg censorship), not feasible (\eg satellite links), or not wanted (\eg privacy concerns). The \ac{ietf} \ac{masque} \ac{wg} was chartered to address these challenges.

    \ac{masque} (see §\ref{sec:masque}) proposes the use of QUIC as a substrate protocol, allowing arbitrary data to be tunneled over QUIC. While this addresses TCP proxy use-cases, it also introduces an alternative layering of \acp{vpns}, where nested reliability can be avoided by leveraging QUIC datagram frames.

    While QUIC and \ac{masque} set out to change our transport protocol usage, the web security model limits browser-based web applications to directly access transport protocol features.
    Protocols like WebSocket (RFC 6455) and WebRTC (RFC 7478) \cite{webrtccommst} were indispensable in rejuvenating static request-response-based web content and benefited from years of deployment of their substrate protocols.
    However, they also inherited their fundamental disadvantages.
    The WebTransport \ac{wg} (see §\ref{sec:webtrans}) addresses these shortcomings by utilizing QUIC as a substrate protocol, exposing its features to browser-based web applications while considering fallback mechanisms to traditional TCP-based connections.
    
    In this article, we present a first look at these most recent transport layer \ac{ietf} standardization efforts \textit{beyond} QUIC \textit{v1}.
    While our work does not cover advances in congestion control schemes such as \ac{bbr} or related standardization work by the \ac{iccrg}, we refer the inclined reader to \cite{recentadvancesintransportlayer} \cite{mlcc}.
    The remainder of this article is structured as follows:
    §\ref{sec:quicv1} briefly introduces QUIC, where §\ref{quic} details future extensions beyond \textit{v1}.
    §\ref{sec:masque} presents recent developments in the usage of QUIC as a substrate protocol within \ac{masque}, where §\ref{sec:webtrans} details the advances of providing novel transport protocol features within the web security model pursued by the WebTransport \ac{wg}.
    Finally, §\ref{sec:interplay} details the interplay of the presented \ac{ietf} standardization efforts, followed by the conclusion in §\ref{sec:conclusion}.

\begin{figure}[t]
    \centering
    \includegraphics[width=1\linewidth]{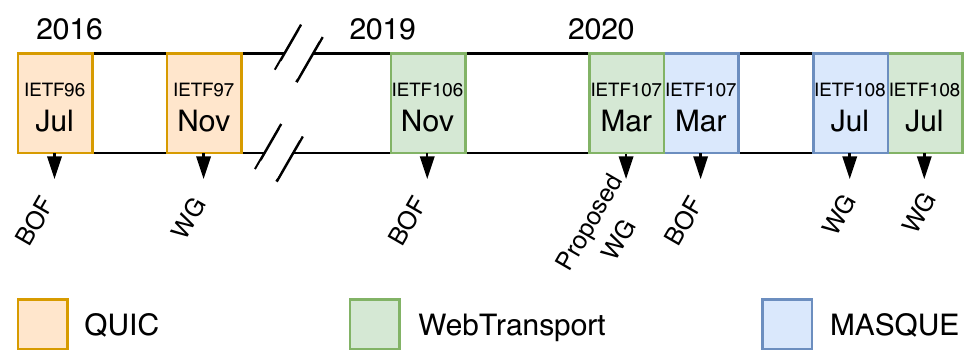}
    \caption[]{\em Timeline of recent IETF transport layer standardization efforts. The QUIC working group was established at \ac{ietf} '97 in 2016, followed by the establishment of \ac{masque} as well as WebTransport at \ac{ietf} '108 in 2020.}
    \label{fig:timeline}
    \vspace{-1em}
\end{figure}

\begin{table*}[t]
    \centering
    \begin{tabularx}{\textwidth}{l|l|c|X}
        \textbf{WG}                     & \textbf{Document}                                             & \textbf{Type}            & \textbf{Reference} \\ \hline
        QUIC                            & \ac{wg} Charter                                               & Charter                  & \href{https://datatracker.ietf.org/wg/quic/about/}{/wg/quic/about/} \\
                                        & An Unreliable Datagram Extension to QUIC                      & WG I-D                   & \href{https://datatracker.ietf.org/doc/draft-ietf-quic-datagram/}{draft-ietf-quic-datagram} \\
                                        & QUIC-LB: Generating Routable QUIC Connection IDs              & WG I-D                   & \href{https://datatracker.ietf.org/doc/draft-ietf-quic-load-balancers/}{draft-ietf-quic-load-balancers} \\
                                        & Compatible Version Negotiation for QUIC                       & WG I-D                   & \href{https://datatracker.ietf.org/doc/draft-ietf-quic-version-negotiation/}{draft-ietf-quic-version-negotiation} \\
                                        & 3GPP Access Traffic Steering Switching and Splitting          & Ind I-D                  & \href{https://datatracker.ietf.org/doc/draft-bonaventure-quic-atsss-overview/}{draft-bonaventure-quic-atsss-overview} \\
                                        & Multipath Extensions for QUIC                                 & Ind I-D                  & \href{https://datatracker.ietf.org/doc/draft-deconinck-quic-multipath/}{draft-deconinck-quic-multipath} \\
& Multipath Extension for QUIC                                  & Ind I-D                  & \href{https://datatracker.ietf.org/doc/draft-liu-multipath-quic/}{draft-liu-multipath-quic} \\
         							    & Greasing the QUIC Bit                                         & Ind I-D                  & \href{https://datatracker.ietf.org/doc/draft-thomson-quic-bit-grease/}{draft-quic-bit-grease} \\
         							    & Sender Control of Acknowledgement Delays in QUIC              & Ind I-D                  & \href{https://datatracker.ietf.org/doc/draft-iyengar-quic-delayed-ack/}{draft-quic-delayed-ack} \\
    \end{tabularx}
    \vspace{2.5pt}
    \caption{Overview of QUIC \ac{ietf} Documents. \textbf{Type} differentiates Document type: WG Charters are denoted as \textbf{Charter}, adopted Internet-Drafts as \textbf{WG I-D}, and individual drafts as \textbf{Ind  I-D}}.
    \label{tab:quicdocs}
    \vspace{-3em}
\end{table*}

\section{QUIC v1} \label{sec:quicv1}

    QUIC was launched by Google in 2012 \cite{quicgoogle} with the goal to provide secure and reliable low latency end-to-end transport.
    Google added support for Chrome in 2013, and by 2017, all Chrome and Android YouTube app users were using QUIC. QUIC provides stream multiplexing without the drawbacks of TCP's \ac{holb}.
    The initial design idea of QUIC was provided by SPDY, which was later standardized as HTTP/2 (RFC 7540), enabling the multiplexing of streams using the same TCP connection.
The \ac{ietf} chartered the QUIC working group in 2016 (see Figure \ref{fig:timeline}) to provide a standards-track specification for a UDP-based, stream-multiplexing, encrypted transport protocol based on Google's pre-standardization implementation and deployment experiences.
    QUIC's \ac{wg} charter holds several goals and milestones relating to the core transport functionality, security, the mapping between different HTTP protocols, the extension of core protocol facilities, and the applicability and manageability of the implications of the protocol.
    QUIC mitigates the \ac{holb} issue by leveraging stream multiplexing in the transport layer, improves on connection establishment times by sending a cryptographic handshake as part of the transport handshake, and provides 1-RTT handshakes for first-time connections as well as 0-RTT handshakes for subsequent connections using TLS 1.3 (RFC 8446).

\section{QUIC Extensions} \label{quic}

    While the working group initially focussed on the web use-case, many QUIC extensions have recently been proposed.
In the following, we will discuss the proposals listed in Table \ref{tab:quicdocs}.

    \textit{An Unreliable Datagram Extension to QUIC:} QUIC transmits a reliable stream of application data where reliability is achieved on a per-stream basis.
    The proposed extension enhances QUIC with the support for unreliable datagrams, aiding in use-cases where reliability is undesired (\eg real-time communication).
    With its reduced handshake latency, unreliable delivery via QUIC  improves on existing solutions such as DTLS (RFC 6347).
    Moreover, its multi-streaming feature can be leveraged to multiplex reliable and unreliable streams within one connection, thereby providing partial reliabilty, and use pluggable congestion control where required.
    Another use-case for unreliable delivery are \acp{vpns}, requiring generic IP packet tunnelling as provided by \ac{masque} (see §\ref{sec:masque}).

    \textit{QUIC-LB: Generating routable QUIC connection IDs:} QUIC maintains a connection ID (CID) per connection, which allows migration during network changes, and provides unlinkability features across connection migration.
    If servers do not provide additional CIDs, they might choose linkable CIDs from servers behind load balancers.
    In this situation, the client either terminates the connection during the migration or remains linkable, violating QUIC's design goal.
QUIC-LB specifies standards for encoding routing information given a small set of configuration parameters.
    Using QUIC-LB, load-balancers communicate the algorithm parameters to generate routable CIDs rather than generating individual CIDs to servers.

    \textit{Compatible version negotiation for QUIC:} Currently, the QUIC server indicates if a client offered version is not accepted, but does not provide information to select a mutually supported version.
    The proposed version negotiation mechanism allows a client and a server to leverage the similarities between different versions and establish a mutually supported/compatible version without the overhead of extra round trips.
    
    \textit{Multipath Extension for QUIC:} The QUIC Multipath extension preserves the single-path QUIC design features while simultaneously using multiple paths for a single connection.
    The introduction and preliminary work on multipath QUIC was presented in \cite{mpquicpaper}.
    Recently, the \ac{3gpp} started a discussion to enable \ac{atsss} service for QUIC on multiple paths, where \ac{ietf} standardization and specifications can be beneficial to attain the \ac{atsss} design goals.
    Table \ref{tab:quicdocs} lists the Informational \ac{3gpp} \ac{atsss} Overview document, as well as multiple individual multipath QUIC drafts which are actively discussed within the \ac{wg}.
    As of now, no consensus has been reached on the adoption of multipath QUIC in general, or a specific proposal in particular.
    While all proposals differ in their way to handle multipath QUIC requirements like linkability between flows, they are commonly in-line with multipath extensions of other transport protocols such as \ac{cmt-sctp} \cite{cmt-sctp-paper} or \ac{mptcp} (RFC 8684) \cite{mpsurvey}.
    These include features like bandwidth aggregation, seamless handovers, and improved user Quality-of-Experience related to the increasing number of multi-homed devices.
    As an example, connection migration can be leveraged on ordinary QUIC connections to move a single QUIC flow from one IP address to another, resulting in a \textit{hard handover}.
    Like \ac{mptcp}, multipath QUIC improves on this, allowing devices to seamlessly switch from one interface to another, thus providing resilience to connection failures.
    Extending on these common multipath features, the primary motivation behind multipath QUIC however lies in the aggregation of all available network resources to send data through a single connection \cite{resourcepooling}.
    While this is useful for \eg large transfers, it also benefits dual-stacked hosts, automatically selecting the best available path in case the quality of the IPv4 and IPv6 paths differ \cite{ippathdiffs}.
    A descriptive example of multipath QUIC is presented in §\ref{sec:interplay}.

    \textit{Greasing the QUIC Bit:} Intermediaries and end-points use the \textit{QUIC Bit} to distinguish QUIC from other protocols.
    A fixed value is currently sent in the QUIC Bit of every packet, thus allowing end-points and intermediaries to depend on a fixed value.
    By leveraging the concept of \textit{GREASE} (Generate Random Extensions And Sustain Extensibility), the \textit{grease\_quic\_bit} transport parameter ensures the future usage of the QUIC Bit by indicating that an end-point is willing to receive QUIC packets regardless of this bit's value.

    \textit{Sender Control of Acknowledgement Delays in QUIC:} A receiver acknowledges the reception of data from the sender. 
    Delaying these acknowledgements reduces the CPU utilization at both sender and receiver and potentially improves throughput.
    However, these benefits are traded off by negatively impacting congestion control and loss recovery.
    The \textit{Sender Control of Acknowledgement Delays in QUIC} extension allows the end-points to advertise the \textit{min\_ack\_delay} transport parameter, which defines the minimum amount of time an ACK can be delayed.

    While these proposals improve on QUIC, there usage requires both communication partners to mutually support an extension.
    As deployment experience with TCP has shown, this can lead to slow adoption, or even no adoption at all. 
    \textit{Pluginzing QUIC} \cite{pquic} enables QUIC end-points to dynamically exchange protocol extensions on a per-connection basis, therefore requiring only one communication partner to feature an extension.

\begin{table*}[t]
    \centering
    \begin{tabularx}{\textwidth}{l|l|c|X}
        \textbf{WG}                     & \textbf{Document}                                             & \textbf{Type}            & \textbf{Reference} \\ \hline
        \ac{masque}                     & \ac{wg} Charter                                               & Charter                  & \href{https://datatracker.ietf.org/wg/masque/about/}{/wg/masque/about/} \\
                                        & Using QUIC Datagrams with HTTP/3                              & WG I-D                   & \href{https://datatracker.ietf.org/doc/draft-ietf-masque-h3-datagram/}{draft-ietf-masque-h3-datagram} \\
                                        & The CONNECT-UDP HTTP Method                                   & WG I-D                   & \href{https://datatracker.ietf.org/doc/draft-ietf-masque-connect-udp/}{draft-ietf-masque-connect-udp} \\
                                        & Requirements for a MASQUE Protocol to Proxy IP Traffic        & WG I-D                   & \href{https://datatracker.ietf.org/doc/draft-ietf-masque-ip-proxy-reqs/}{draft-ietf-masque-ip-proxy-reqs} \\
                                        & The CONNECT-IP method for proxying IP traffic                 & Ind I-D                  & \href{https://datatracker.ietf.org/doc/draft-kuehlewind-masque-connect-ip/}{draft-kuehlewind-masque-connect-ip} \\     
                                        & QUIC-Aware Proxying Using CONNECT-UDP                         & Ind I-D                  & \href{https://datatracker.ietf.org/doc/draft-pauly-masque-quic-proxy/}{draft-pauly-masque-quic-proxy} \\                                         
                                        & Discovery Mechanisms for QUIC-based Proxy Services            & Ind I-D                  & \href{https://datatracker.ietf.org/doc/draft-kuehlewind-masque-proxy-discovery/}{draft-kuehlewind-masque-proxy-discovery} \\
                                        & Transport Considerations for IP and UDP Proxying              & Ind I-D                  & \href{https://datatracker.ietf.org/doc/draft-westerlund-masque-transport-issues/}{draft-westerlund-masque-transport-issues} \\

        \rowcolor{gray!100}
        WebTransport                    & \ac{wg} Charter                                               & Charter                  & \href{https://datatracker.ietf.org/wg/webtrans/about/}{/wg/webtrans/about/} \\
        \rowcolor{gray!100}
                                        & The WebTransport Protocol Framework                           & WG I-D                   & \href{https://datatracker.ietf.org/doc/draft-ietf-webtrans-overview/}{draft-ietf-webtrans-overview} \\
        \rowcolor{gray!100}
                                        & WebTransport using HTTP/2                                     & Ind I-D                  & \href{https://datatracker.ietf.org/doc/draft-kinnear-webtransport-http2/}{draft-kinnear-webtransport-http2} \\
        \rowcolor{gray!100}
                                        & WebTransport over HTTP/3                                      & Ind I-D                  & \href{https://datatracker.ietf.org/doc/draft-vvv-webtransport-http3/}{draft-vvv-webtransport-http3} \\
        \rowcolor{gray!100}
                                        & WebTransport over QUIC                                        & Ind I-D                  & \href{https://datatracker.ietf.org/doc/draft-vvv-webtransport-quic/}{draft-vvv-webtransport-quic}
    \end{tabularx}
    \vspace{2.5pt}
    \caption{Overview of MASQUE and WebTransport \ac{ietf} Documents. \textbf{Type} differentiates Document type: WG Charters are denoted as \textbf{Charter}, adopted Internet-Drafts as \textbf{WG I-D}, and individual drafts as \textbf{Ind  I-D}}.
    \label{tab:docs}
    \vspace{-3em}
\end{table*}

\section{MASQUE}
\label{sec:masque}

    Driven by the shortcomings of proxying mechanisms like native HTTP Proxies (unencrypted, HTTP/TCP), \ac{socks} (unencrypted signaling, TCP and UDP), HTTP CONNECT (encryption optional, TCP), or transparent TCP Proxies (must be on-path, mandatory to use, TCP), the IETF \ac{masque} \ac{wg} (see Figure \ref{fig:timeline}) was formed in early 2020.
    \ac{masque} is chartered to develop mechanisms that will allow arbitrary connections to be tunneled within a single HTTP/3 connection using explicit client-initiated signaling.
    Besides the existing request/response model and authentication mechanisms of HTTP, which can be leveraged for service and parameter negotiation, QUIC's unified congestion controller will greatly improve on the uncoupled flows handled by traditional proxies, and allow multiple client-initiated reliable and unreliable connections to be transported within a single HTTP/3 connection.
    To address censorship use-cases, the tunneled data will be indistinguishable to arbitrary encrypted HTTP connections on the wire, preventing hints which possibly expose the nature of the connection to adversaries.
    Moreover, to address instances where UDP and/or HTTP/3 is actively blocked on the client-proxy leg of the connection, the \ac{masque} \ac{wg} will consider HTTPS/TCP as a fallback.

    Initially proposed within the QUIC \ac{wg}, \textit{Using QUIC Datagrams with HTTP/3} (see Table \ref{tab:docs}) was recently moved to and adopted by the \ac{masque} \ac{wg} as a WG Internet draft.
    While the unreliable datagram extension of QUIC (see §\ref{quic}) provides a mechanism to send reliable and unreliable data simultaneously leveraging the security and congestion-control properties of QUIC, it is unable to de-multiplex application contexts.
    \textit{Using QUIC Datagrams with HTTP/3} adds flow identifiers for HTTP/3 applications at the start of the frame payload. This \textit{Datagram-Flow-Ids} represent bidirectional flows in a single QUIC connection and allow multiplexing and de-multiplexing of the application data.
    This concept is leveraged within \ac{masque} as well WebTransport (see §\ref{sec:webtrans}).

    As a primary focus for the \ac{wg}, \textit{CONNECT-UDP} (see Table \ref{tab:docs}) proposes a UDP-based counterpart to the TCP-only \textit{HTTP CONNECT} method.
While it would be possible to reuse \textit{HTTP CONNECT} for UDP, existing implementations would fallback to TCP on the proxy-server leg of the connection, which should be avoided.
    However, \textit{CONNECT-UDP} will be supported on HTTP/1.1, 2, and 3, and therefore provides a TCP fallback mechanism on the client-proxy leg of the connection as detailed earlier.
    Using the \textit{CONNECT-UDP} header, the client instructs the proxy to open a UDP connection to a provided URI.
    For HTTP/3, QUIC datagram frames are leveraged, providing a proxied unreliable connection between client and server.
    This enables connections to multiple servers to be transported within the same client-proxy HTTP/3 connection, which are multiplexed and de-multiplexed using \textit{Datagram-Flow-Ids}
    While the chaining of multiple proxies is supported, a proxy receiving \textit{CONNECT-UDP} can either connect to the indicated target or to an upstream proxy.
    To use UDP on an end-to-end path, all involved proxies have to support HTTP/3 leveraging QUIC datagram frames.
    Following successful negotiation, all intermediaries will switch to tunnel mode and restrict to forwarding packets until the connection is closed. 

    Besides \textit{CONNECT-UDP}, the requirements for generic \textit{IP Proxying} (see Table \ref{tab:docs}) addressing traditional VPN use-cases are actively discussed, and were recently adopted as a WG Internet draft.
Favoring HTTP/3 using QUIC datagram frames to prevent nested reliability, a fallback to HTTP/2 is also supported, leveraging both protocols multiplexing capabilities to run multiple IP proxied connections over the same HTTP connection.
    For this purpose, an IPv4 or IPv6 session has to be established between the end-points, including support for IP address assignment requests, route negotiation, and client and server identification as well as authentication.
    Where \textit{IP Proxying} lays out the requirements for proxying IP packets, \textit{CONNECT-IP} (see Table \ref{tab:docs}) proposes a specific method to enable IP proxying using HTTP/3 connections, thus partially covering the outlaid requirements. 
    A descriptive use-case of \textit{IP Proxying} is presented in §\ref{sec:interplay}.

    To proxy arbitrary QUIC connections, \textit{QUIC-Aware Proxying Using CONNECT-UDP} (see Table \ref{tab:docs}) addresses the specifics of tunneling QUIC over QUIC for long header packets, \eg the encapsulation and encryption overhead of nested QUIC connections, as well as the forwarding of short header QUIC packets on established connections by leveraging connection IDs.

    Exceeding the presented efforts, supplemental topics are discussed within the \ac{wg} which are also shown in Table \ref{tab:docs}.
    \textit{Discovery Mechanisms for QUIC-based Proxy Services} discusses mechanisms to enable clients to be able to discover non-transparent \ac{masque} proxies, while \textit{Transport Considerations for IP and UDP Proxying in MASQUE} addresses challenges to preserve end-to-end properties of the proxied flows.

\section{WebTransport}
\label{sec:webtrans}

    The web security model shapes the Internet landscape.
    While abstracting transport protocol features to application layer protocols and exposing them to web developers, browser-based web applications became truly interactive and highly dynamic, radically replacing static request-response based content.

    The TCP streams exposed by the WebSocket protocol (RFC 6455) provide bidirectional ordered delivery and suffer from \ac{holb} as well as mandatory reliability, making it a poor fit for real-time communication or latency-sensitive applications.
    This is improved by bootstrapping WebSocket onto HTTP/2 (RFC 8441), which multiplexes arbitrary streams in a single HTTP/2 connection, hence eliminating HTTP \ac{holb}, but still suffering from TCP \ac{holb}.
    Layering WebSocket onto HTTP/3 would solve this issue.
However, existing disadvantages persist, requiring additional round-trips for the WebSocket protocol handshakes for every stream, limiting connection initiation to clients only, and lacking support for unreliable transport.

    WebRTC (RFC 7478) \cite{webrtccommst} data channels improve on this while leveraging \ac{sctp} (RFC 4960), providing ordered and unordered delivery, partial reliability, and eliminating \ac{holb}.
As \ac{sctp} faced deployment challenges (see §\ref{sec:intro}), \ac{sctp} WebRTC data channels use UDP as a substrate, a pattern also embraced by QUIC (see §\ref{sec:quicv1}).

    The IETF WebTransport \ac{wg} (see Figure \ref{fig:timeline}) was formed to provide a mapping of HTTP and QUIC-based protocols to a web interface API developed by the \ac{w3c} \cite{w3cwebtrans} honoring the web security model.
    The utilized protocols (referred to as \textit{transports}) mandate uni- and bi-directional streams, datagram support, and encryption.
    Moreover, optional properties are defined, which rely on features of specific protocols.
    These include stream independence to prevent \ac{holb}, partial reliability to prevent retransmissions of datagrams, pooling support to share a unified congestion controller, connection migration to keep connections alive if the path changes, and bandwidth prediction to aid use-cases like video streaming or real-time gaming.

    While the core incentives of WebTransport have been discussed since 2018 as QUIC standardization progressed, the \ac{wg} was chartered in March 2020, currently defining the requirements of WebTransport, and the requirements on the utilized \textit{transports} \textit{Http2Transport}, \textit{Http3Transport}, as well as \textit{QuicTransport} (see Table \ref{tab:docs}).
    A descriptive example of WebTransport is presented in §\ref{sec:interplay}.

    \textit{Http2Transport} allows WebTransport peers to multiplex arbitrary bidirectional streams over HTTP/2 connections, where either end-point can initiate a new stream.
    While WebTransport and regular HTTP data can be multiplexed simultaneously, intermediaries traversed must explicitly support WebTransport.
    Additionally, TCP \ac{holb} remains an issue, and the mandated support for unidirectional streams and unreliable delivery are noticeably missing.
    While unidirectional streams can be forged by requiring bidirectional streams to only use one half of the connection, unreliability can not be provided as TCP forcibly retransmits HTTP/2.
    As datagrams are \textit{not} expected to be reliably delivered, but they \textit{might} if the \textit{transport} is using a TCP-based protocol, the specification also covers this fallback case.
    Additionally, \textit{Http2Transport} does feature pooling support, which ensures that a shared congestion controller between multiple \textit{transports} sharing the same HTTP connection can be used.

    \textit{Http3Transport} does support all requirements covered by \textit{Http2Transport}, and extends on it by also providing unidirectional streams, unreliable delivery leveraging QUIC datagram frames with HTTP/3 (see §\ref{sec:masque}), as well as stream independence which eliminates \ac{holb}.

    Lastly, \textit{QuicTransport} offers a minimal protocol on top of QUIC, where WebTransport concepts are directly mapped to the corresponding QUIC counterparts if applicable.
    The main design goal is a low overhead protocol, minimizing implementation effort and complexity for extending existing QUIC stacks with \textit{QuicTransport} capabilities.
    \textit{QuicTransport} satisfies all WebTransport design requirements except pooling support.
    
    Besides the three presented proposals, a fourth option of \textit{FallbackTransport} (no active document) is discussed within the \ac{wg}.
    Aiming at a mapping to HTTP/1.1 and HTTP/2, multiplexed streams can be simulated on top of the WebSocket protocol, where the existing standardized WebSocket mappings to the HTTP protocols are utilized as-is.

While \textit{QuicTransport} offers a solution with low overhead, low complexity, and minimal implementation effort, \textit{Http3Transport} offers pooling support as well as HTTP features like status codes, headers, load balancing, and rerouting, possibly outweighing the increased complexity and interdependency.
    Acknowledging these advantages, an adoption call was recently issued for the \textit{Http3Transport} proposal, aiming to focus the \ac{wg}s resources at WebTransport over HTTP/3 in the foreseeable future.

\begin{figure}[t]
    \centering
    \includegraphics[width=1\linewidth]{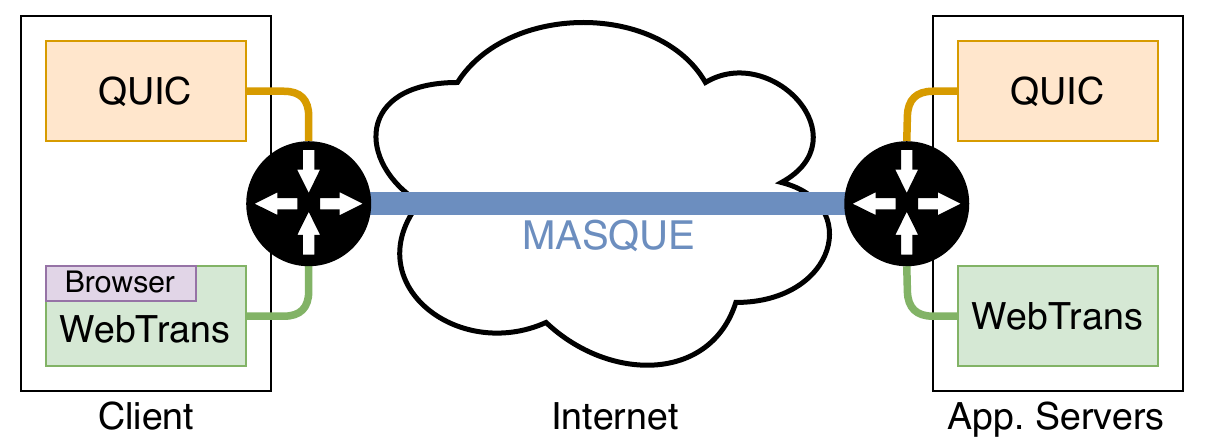}
    \caption[]{\em Remote office use-case: A client requests resources via QUIC and WebTransport from Application Servers, which are multiplexed over a \ac{masque} tunnel proxying arbitrary IP packets.}
    \label{fig:interplay}
    \vspace{-1em}
\end{figure}

\section{Interplay} \label{sec:interplay}

\begin{figure}[t]
        \centering
        \includegraphics[width=1\linewidth]{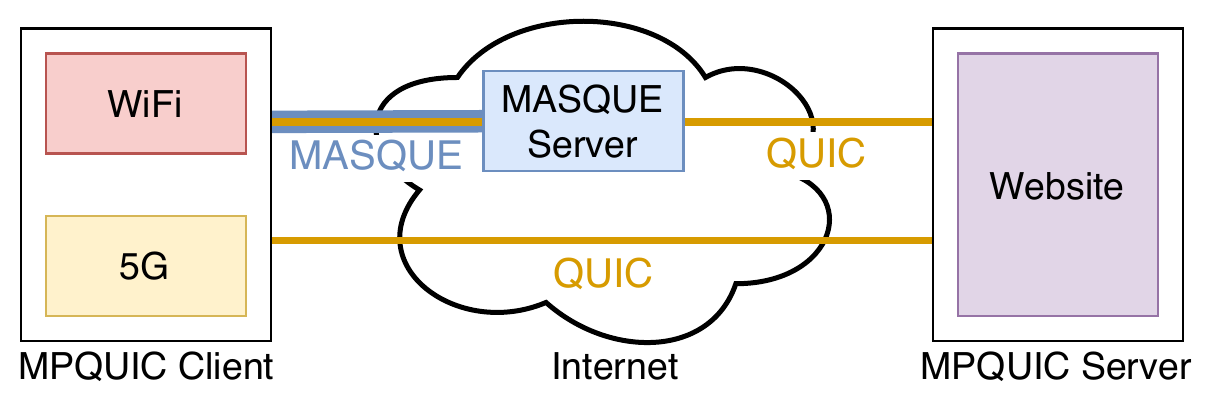}
        \caption[]{\em Mobile use-case: A client is connected to a server using multipath QUIC (MPQUIC), where one QUIC connection is proxied via a \ac{masque} server.}
        \label{fig:mpquic}
        \vspace{-1em}
    \end{figure}

    While all discussed protocols and extensions propose vastly different approaches, their interplay extends on QUIC's take to address transport layer ossification challenges.
    We present two different use-cases highlighting their combined benefits.
    Figure \ref{fig:interplay} showcases a remote office scenario.
    A client requests resources using plain QUIC as well as WebTransport via a browser.
    The client runs a VPN service leveraging a \ac{masque} tunnel to proxy arbitrary IP packets, connecting to a VPN gateway at the main office that de-multiplexes the tunnel and proxies the requests to their respective application servers.
    A mobile use-case is presented in Figure \ref{fig:mpquic}.
    A multipath QUIC client is connected to a multipath QUIC server using WiFi and 5G simultaneously, thus featuring multiple end-to-end paths.
    One connection is proxied using a \ac{masque} server, where the end-to-end QUIC connection is tunneled within the \ac{masque} QUIC connection.
    Benefitting from both the proxied \ac{masque} connection optimized for the access network as well as the multipath capabilities, the client's packet scheduler can dynamically select the optimal path and seamlessly re-route packets in case of path property changes or connection losses.

\section{Conclusion}
\label{sec:conclusion}

    The transport layer is evolving.
    With QUIC at the core of this renewal, its future versions will build on the foundation of QUIC v1 deployed on the Internet, thereby extending its reach to increasingly more application areas.
    While multiple extensions improve on QUIC itself, \ac{masque} shows promise to supersede traditional proxies and \acp{vpns}, and WebTransport will further enhance the rejuvenation of the Web, thus aiding the development of next-generation Web applications.
    While this article offered a first impression at the recent transport layer \ac{ietf} standardization efforts \textit{beyond} QUIC \textit{v1}, the presented protocols and extensions propose different approaches to address long-standing problems, and their interplay extends on QUIC's take to address ossification challenges. 
    Marking only the beginning, a new era of protocols is about to emerge.

\section*{Acknowledgments} \label{sec:acks}
We would like to thank Lars Eggert and Mirja Kühlewind for their feedback and guidance, as well as the editor and the reviewers for their valuable remarks.

\bibliographystyle{IEEEtran}

\vspace{-4em}

\begin{IEEEbiographynophoto}{Mike Kosek}
    is a PhD Student at TUM, Germany. His current research focuses on Internet architectures in general, and transport protocol standardization, development, and deployment, in particular.
\end{IEEEbiographynophoto}

\vspace{-4em}

\begin{IEEEbiographynophoto}{Tanya Shreedhar}
    is a PhD student at IIIT-Delhi, India. Her research interests include next-generation networks and systems with a focus on transport layer protocols.
\end{IEEEbiographynophoto}

\vspace{-4em}

\begin{IEEEbiographynophoto}{Vaibhav Bajpai}
    is a senior researcher at TUM, Germany. He is interested in performance and management of next-generation networked systems.
\end{IEEEbiographynophoto}

\end{document}